\documentclass{JINST}
 
\usepackage{graphicx}
\usepackage{caption}
\usepackage{subcaption}
\usepackage{sidecap}

\usepackage{procpaperdefs}

\title{ A Leakage Current-based Measurement of the Radiation Damage in
        the ATLAS Pixel Detector }

\author{Igor~V.~Gorelov~on~behalf~of~the~ATLAS~Collaboration\\
\llap{$^a$}Department of Physics and Astronomy, University of New Mexico,\\
           Albuquerque, NM 87131, USA.\\
E-mail: \email{igor.gorelov@cern.ch}}

\abstract{%
  A measurement has been made of the radiation damage incurred by the
  ATLAS Pixel Detector barrel silicon modules from the beginning of
  operations through the end of 2012.  This translates to hadronic
  fluence received over the full period of operation at energies up to
  and including \(8\,{\rm TeV}\).  The measurement is based on a per-module
  measurement of the silicon sensor leakage current.  The results are
  presented as a function of integrated luminosity and compared to
  predictions by the Hamburg Model.  This information can be used to
  predict limits on the lifetime of the Pixel Detector due to current,
  for various operating scenarios.
}

\keywords{Si microstrip and pad detectors; Pixelated detectors; Radiation-hard detectors}

\begin{document}
%
%
\section{Introduction}
\label{sec:intro}
  The ATLAS Pixel Detector~\cite{Aad:2008zz} is constructed from
  planar silicon pixel modules arranged in a barrel and disk geometry.
  It is centered on the interaction point at the heart of the ATLAS
  Detector~\cite{Aad:2008zzm} at the Large Hadron Collider
  (LHC).  A linear relationship between leakage current and a hadronic particle 
  fluence is expected to apply to the pixel silicon sensors and is given
  by 
 \begin{equation}
   \Delta{I} = \alpha\cdot\flue\cdot{V},
 \label{eq:i-phi}
 \end{equation}
  where \(\Delta{I} \) is the difference in leakage current at fluence
  \( \flue\) relative to the value before irradiation of the
  physical volume \( V \), and \( \alpha \) is the current-related
  damage coefficient~\cite{Moll:1999nh}.  The goal of this study is to
  measure the leakage currents of a representative sample of sensors in
  the ATLAS Pixel Detector in order to monitor and understand the sensor
  damage resulting from increasing radiation dose.  The measured results
  are compared to predictions based on the Hamburg
  Model~\cite{damage-prediction} in which reverse-bias current is
  generated by the introduction of defect levels in the band gap due to
  non-ionizing energy loss (NIEL) of through-going particles.  The defects
  migrate and recombine through several processes that result in both
  short-term beneficial annealing (which reduces the current) and
  long-term anti-annealing (which increases the current even in the
  absence of further radiation).  The model can be calibrated to the
  data to predict the lifetime of the sensors and the Pixel Detector as
  a whole for various operating scenarios.
\par 
  The detector is instrumented as three barrels, specifically the
  Layer-0 at \(r=50.5\mm\), the Layer-1 at \(r=88.5\mm\), and the Layer-2
  at \(r=122.5\mm\), comprising in total 1456 modules.
  The endcap areas are instrumented with three disks located at
  \(z=\pm495\mm,\,\pm580\mm \) and \(\pm650\mm\) in each of the forward
  and backward regions, comprising an additional 288 modules.  The Pixel
  Detector and the other elements of the ATLAS Inner Detector span a
  pseudorapidity range 
  \(\rapid< 2.5\)~\footnote{ATLAS uses a right-handed coordinate
  system with its origin at the nominal interaction point (IP) in the
  centre of the detector and the $z$-axis along the beam pipe. The
  $x$-axis points from the IP to the centre of the LHC ring, and the
  $y$-axis points upward. Cylindrical coordinates $(r,\phi)$ are used
  in the transverse plane, $\phi$ being the azimuthal angle around the
  beam pipe. The pseudorapidity is defined in terms of the polar angle
  $\theta$ as $\eta=-\ln\tan(\theta/2)$.}.
  The pixel sensor modules are mounted on mechanical/cooling supports,
  called staves, in the barrel region.  Thirteen modules are mounted on
  a stave with identical layouts for all layers.  The temperature is
  maintained stable by an evaporative cooling system.
  The entire \(\sim\!{1.7}{\m}^{2}\) sensitive area of the ATLAS pixel
  detector is covered with 1744 identical modules.  Each module has an
  active surface of \(6.08\times1.64\cma\).
  The details of the pixel sensor geometry and layout can be found 
  in~\cite{Aad:2008zz}.
\par 
  The radiation field in the region of the detector has been predicted
  with the aid of a {\sc FLUKA} software package~\cite{Baranov:2005ewa}.
  The {\sc FLUKA} transport code to cascade the particles through the
  ATLAS detector material has been used together with {\sc PYTHIA~6.2}
  or {\sc PHOJET} generating proton-proton minimum-bias events as an
  input to {\sc FLUKA}~\cite{Aad:2008zz}.
\par 
  We assume that the dominant radiation damage type is displacement
  defects in the bulk of the pixel sensor, initiated by hadronic species
  and caused by NIEL.  Surface ionization is
  neglected in this treatment.  Charged pions are expected to dominate
  the bulk damage for the radii covered by the Pixel Detector.  Albedo
  neutrons originating in the outer ATLAS detectors also contribute.
  The resulting displacement defects increase the reverse leakage
  current, degrade the charge collection efficiency, and change the
  effective doping concentration which directly determines the depletion
  voltage.  We define the effective fluence \flue as the number of
  particles causing damage equivalent to that of \( 1\!\MeV \) neutrons
  traversing \( 1\cma \) of a sensor's surface.  The fluence \( \flue \)
  accumulated by ATLAS pixel detector, and measured in units
  of \({\cm}^{-2}\), is expected to be proportional to integrated
  luminosity \( \IntL \), measured in \( \invpb \).
\section{Measured Quantities}
  The parameter \(\alpha\) in Equation~\ref{eq:i-phi} has been measured
  under a variety of conditions, for example \cite{Moll:1999nh}.  The
  linear form applies to the leakage currents drawn by sensors past
  their beneficial annealing periods.  At the beginning of data-taking
  and during beneficial annealing periods, the sensors draw currents at
  levels comparable to dark currents;
  thus any hardware implementations for leakage current measurement
  should provide access to the lowest current range possible.
\par 
  The reverse bulk generation current depends on the sensor temperature
  $T$.  The current measurements \( I(T) \) are normalized to the
  reference temperature \(T_{R}\) (\(0\degc\))
  with 
%
%
  \begin{equation}
    I(T) = I(T_{R})/R(T),\,\, {\rm where} \,\, 
    R(T) = ( T_{R}/T )^{2}\cdot\exp{\left( {-\frac{E_g}{2k_{B}}}(1/T_{R}-1/T) \right), }
  \label{eq:i-temp}
  \end{equation}
  where \( E_{g} = 1.21\ev \) is the energy of the silicon band gap,
  \(k_{B} \) is the Boltzmann constant, and \(T\,{\rm and}\,T_{R} \) are
  measured in \degk~\cite{Chilingarov}.
\par
\begin{figure}[htbp] 
    \begin{center}
      \includegraphics[width=1.0\textwidth]{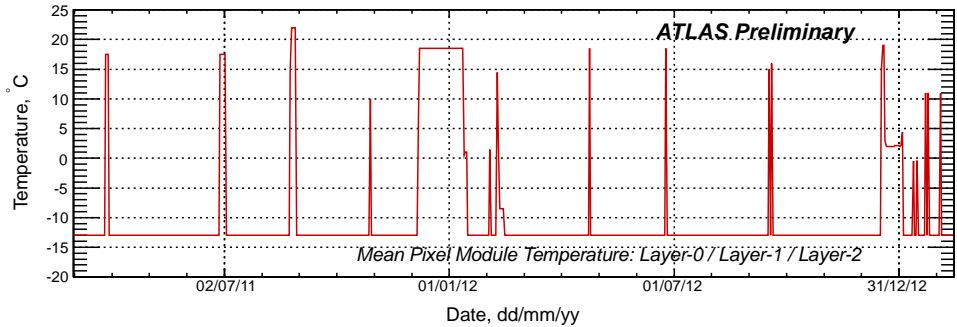}
    \end{center}
    \caption{  The mean temperature (\degc) of all barrel pixel modules, 
               versus time (days)~\cite{ATL-INDET-PUB-2014-004}.}
    \label{temperature-profile}
  \end{figure}

  The mean pixel module temperature for all barrel pixel modules is
  shown versus time in
  Figure~\ref{temperature-profile}~\cite{ATL-INDET-PUB-2014-004}.
  There were a few exceptions when the temperature rose to ambient due
  to cooling interruptions and calibration scans.
  Temperature is read out from every module continuously and averaged
  over 30-minute intervals.  Temperature data (as well as other slow
  control data) are committed into the ATLAS Detector Control System
  (DCS) database.
\section{The ATLAS Current Measurement Subsystem}
  The ATLAS High Voltage Patch Panel 4 (HVPP4) serves as a fan-out point
  for the bias voltages delivered by Type II boards from
  Iseg~\cite{Iseg-specs}
%
%
  high voltage power supplies to the pixel modules.  The leakage current
  is monitored at the pixel module granularity level by a Current
  Measurement Board (CMB) system.  A CMB
  is mounted on every Type~II fan-out board. The bias voltage to the
  sensors is provided by the Iseg channels.
\par
\begin{figure}[htbp] 
    \begin{center}
      \includegraphics[width=0.50\textwidth]{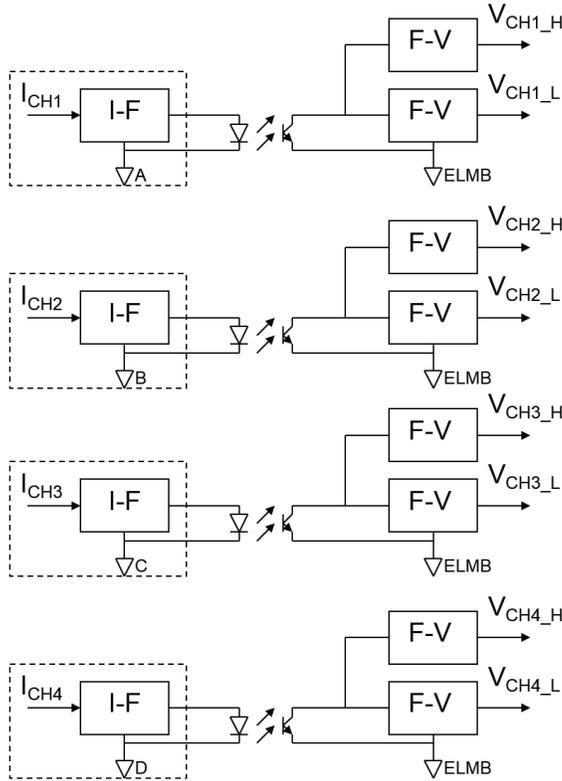}
    \end{center}
    \caption{The Current Measurement Board circuit}.
    \label{CMB-circuit} 
  \end{figure}

 
  In the present data-taking era, when the radiation damage of the
  sensors is relatively low, 6 or 7 pixel modules are fed by one Iseg
  power supply channel.  At some point after inversion of the sensors,
  before the current drawn by 6 or 7 pixel modules reaches the Iseg
  limit, power supplies can be added until the system provides one Iseg
  power supply channel per pair of pixel modules.  The CMB system allows
  simultaneous measurement of the leakage currents in 4 pixel modules in
  the current range from \(0.04\mkamp\) to \(2\mamp\) 
  %
  %
  with a precision of better than \(20\,\%\) per module.  ATLAS uses
  \(21\) CMBs in Layer-0, \(16\) in Layer-1, and \(16\) in Layer-2.  The
  analog current measurements are digitized by the \(64\)-channel ATLAS
  Embedded Local Monitoring (ELMB) board and sent via the Controller Area
  Network (CAN) bus to the DCS database (see~\cite{Aad:2008zz} for
  details on the Type~II fan-out board, ELMB, CAN, and DCS). The building
  block of the CMB circuit is a current to frequency converter optically
  coupled to a frequency to voltage converter as shown in
  Figure~\ref{CMB-circuit}.  Each CMB holds \(4\times2\) current measurement
  circuits and provides the current measurement for \(4\) pixel modules.
  Two similar circuits with different gains are allocated per CMB
  channel serving the same module.  The channels are isolated from each other
  and from the pixel module readout system.
%
%
  For each pixel module, the leakage current range
  \([10^{-8}-10^{-5}]\amp\) is covered by the high-gain channel, and the
  range \([10^{-6}-2\cdot10^{-3}]\amp\) is covered by the low-gain
  channel.  The output CMB voltage range is defined by one of five
  available ELMB input voltage ranges, which are \([0-25]\mvolt\),
  \([0-100]\mvolt\), \([0-1]\volt\), \([0-2.5]\volt\), and
  \([0-5]\volt\).  The output range \([0-5]\volt\) is necessary to
  comply with the specifications of the digital ELMB board.  The 16-bit
  ADC of the ELMB provides a resolution of (ELMB
  range)/\(({2}^{16}-1)\).  The range \([0-1]\volt\) was used for data
  taken through 2011; the range was changed to \([0-5]\volt\) in 2012 as
  the high-gain channels reached saturation, corresponding to pixel
  leakage currents greater than \(10^{-5}\amp\).  After LHC shutdown of
  2013-2014, the data-taking will resume with the low-gain channels.
  Before installation, each CMB is calibrated offline assuming a linear
  model.  The calibrated gains of the channels are stored in the
  database.
  The pedestals are re-calibrated {\it in situ}.
\section{Current Measurements, 2011-2012}
  Figure~\ref{cmbpaper-lumi} shows the leakage current recorded by the
  CMB system from the beginning of ATLAS data-taking through the end of
  2012, as a function of integrated luminosity, for modules in all three
  Pixel Detector barrel layers~\cite{ATL-INDET-PUB-2014-004}.
  Installation of the CMB system proceeded concurrently with early
  data-taking, and this is the reason that some Layer-2 data are missing
  from the early part of the run. It is foreseen that the disks will be
  instrumented with current monitor boards prior to the start of
  data-taking in 2015 (Run 2).  
  The points at which LHC and cooling were off, hence beneficial
  annealing of the silicon was unsuppressed, are apparent as
  discontinuities in this figure.  This effect, producing the sawtooth
  features, is due to the increase of the effective doping concentration
  of the silicon due to annealing 
  of acceptors~\cite{Moll:1999nh, damage-prediction}.
\par 
  Figure~\ref{cmbpaper-lumi} also includes a
  prediction~\cite{damage-prediction} of the leakage current for the
  detailed ATLAS geometry and LHC luminosity profile based on the
  Hamburg Model. The predictions of the fluence \(\flue\) according to
  the luminosity profile are made using the {\sc FLUKA} transport
  code~\cite{Baranov:2005ewa}. The predictions are uniform w.r.t. the
  azimuthal angle \(\phi \) and suppose a forward-backward symmetry
  w.r.t. the polar angle \(\theta \), i.e. \(\flue(\eta)=\flue(-\eta)
  \).
  At the end of Run 1 with \(\IntL=30\ifb \), the Layer-0 modules have
  received a total fluence 
  \(\flue\eqsim7.0\cdot10^{13}{{\rm \,cm}^{-2}}\), 
  while Layer-1 and Layer-2 have been irradiated to
  \(\flue\eqsim3.0\cdot10^{13}{{\rm \,cm}^{-2}}\) and
  \(\flue\eqsim1.8\cdot10^{13}{{\rm \,cm}^{-2}}\) respectively.
  The uncertainty of the leakage current predictions comprises the
  uncertainty of the calculations of the hadronic particle spectra
  initiated by \(\propro\) interactions in the ATLAS detector and the
  uncertainty of the Hamburg model in describing radiation damage effects
  inflicted by NIEL of hadrons crossing the volume of the
  silicon pixel sensors.
\begin{SCfigure}
  \centering
   \captionof{figure}{ ATLAS Pixel module leakage current versus integrated LHC
              luminosity~\cite{ATL-INDET-PUB-2014-004}. 
              The currents are averaged over their layer for
              all modules equipped with Current Measurement Boards
              within the layer.  The current is continuously monitored
              by the ATLAS Detector Control System.  A prediction based
              on the Hamburg Model is included.  Discontinuities are due
              to annealing during LHC and cooling stops.} 
  \includegraphics[width=0.5\linewidth]
  {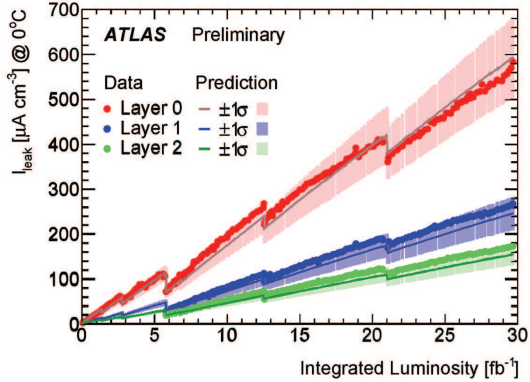}
  \label{cmbpaper-lumi}
\end{SCfigure}
  Figure~\ref{fig:phi-map} shows the distribution in azimuthal angle of
  modules whose current is sampled by CMBs in the ATLAS Pixel
  Detector.
  Figure~\ref{fig:phi-current} shows the data for each of four quadrants
  in the azimuthal angle as they are defined by
  Figure~\ref{fig:phi-map}~\cite{ATL-INDET-PUB-2014-004}.
  No systematic difference in a module leakage current is observed
  w.r.t. to the quadrant where the module is located.
\begin{figure}[htbp]
    \centering
    \begin{subfigure}[t]{.4\textwidth}
        \centering
        \includegraphics[width=1.05\linewidth]{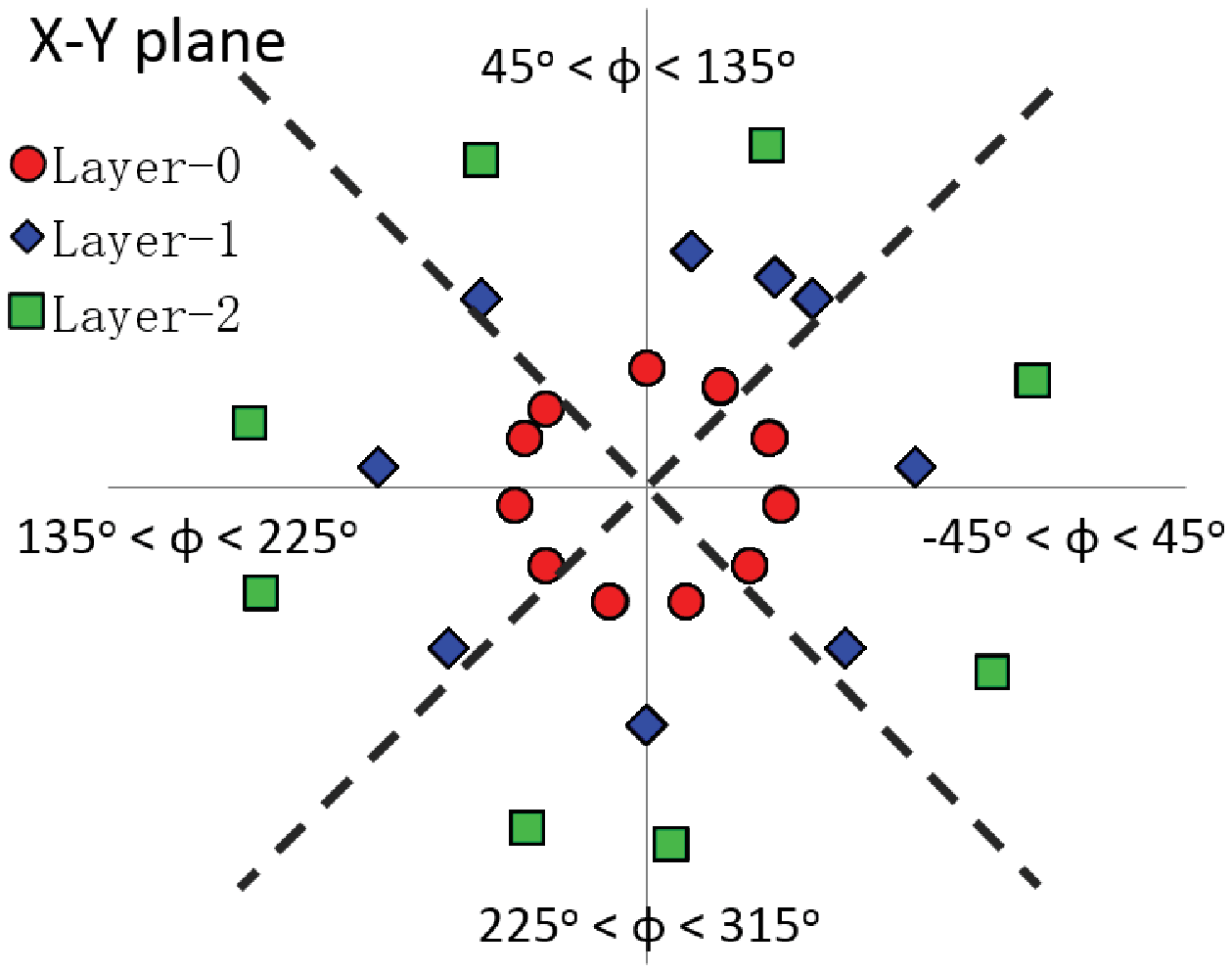}
        \caption{ A diagramm of modules.}
        \label{fig:phi-map}
    \end{subfigure}%
    \begin{subfigure}[t]{.6\textwidth}
        \centering
        \includegraphics[width=1.0\linewidth]{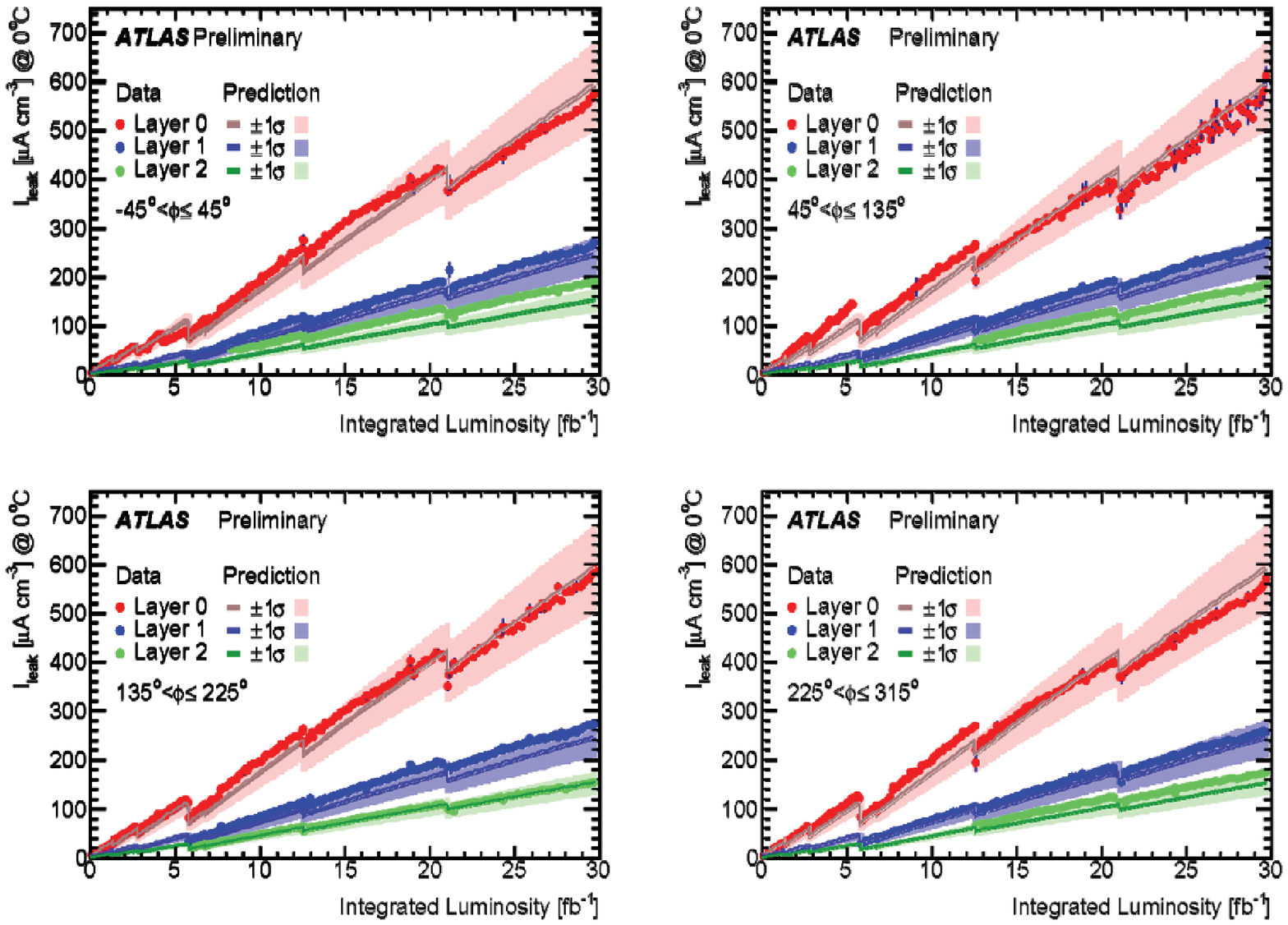}
        \caption{ Module leakage current versus integrated LHC luminosity.}
         \label{fig:phi-current}
    \end{subfigure}
    \caption{(a)~Sampling of 4 quadrants in azimuthal angle with modules
      instrumented by CMBs.~(b)~The module leakage current in each of the 4 
      quadrants in azimuthal angle~\cite{ATL-INDET-PUB-2014-004}.  The
      currents are averaged over each layer for all modules equipped
      with Current Measurement Boards within the particular quadrant of
      the layer. Predictions based on the Hamburg Model are included.}
    \label{cmbpaper-phi}
\end{figure}
  Figure~\ref{cmbpaper-eta-diagram} shows the instrumented module
  distribution, segmented into five sectors in pseudorapidity. The
  pseudorapidity acceptance of each sector of each layer is indicated.
  Plots in Figure~\ref{fig:eta1} through Figure~\ref{fig:eta5}, show the
  leakage current data for each of the 5 ranges in
  pseudorapidity~\cite{ATL-INDET-PUB-2014-004}. 
  Figure~\ref{fig:eta6} shows the current data versus pseudorapidity
  for the sectors defined by Figure~\ref{cmbpaper-eta-diagram}, for
  integrated luminosity 25\invfb~\cite{ATL-INDET-PUB-2014-004}.  The
  possible slight asymmetry of the data points in Figure~\ref{fig:eta6}
  is thought to be due to the asymmetric beam profile.  Yet the
  distribution is in agreement within the uncertainties with the leakage
  current predictions which are symmetrical according to the fluence
  simulation~\cite{Baranov:2005ewa}.
\begin{figure}
\centering
  \begin{minipage}{.66\textwidth}
  \centering
    \includegraphics[width=1.0\linewidth]{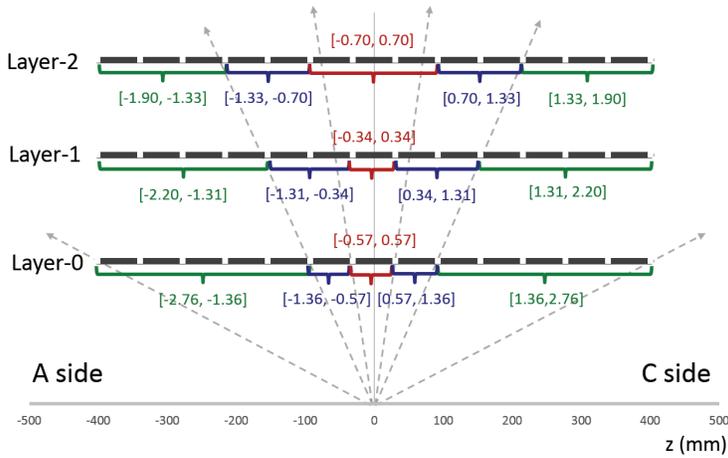}
  \end{minipage}%
  \begin{minipage}{.33\textwidth}
  \centering
    \captionof{figure}{ A diagram of the distribution of instrumented modules in
                        each of five pseudorapidity ranges. }
  \label{cmbpaper-eta-diagram}              
  \end{minipage}
\end{figure}
  \begin{figure}[htbp] 
   \centering
      \begin{subfigure}{0.32\textwidth}
      \centering
        \includegraphics[width=1.0\linewidth]{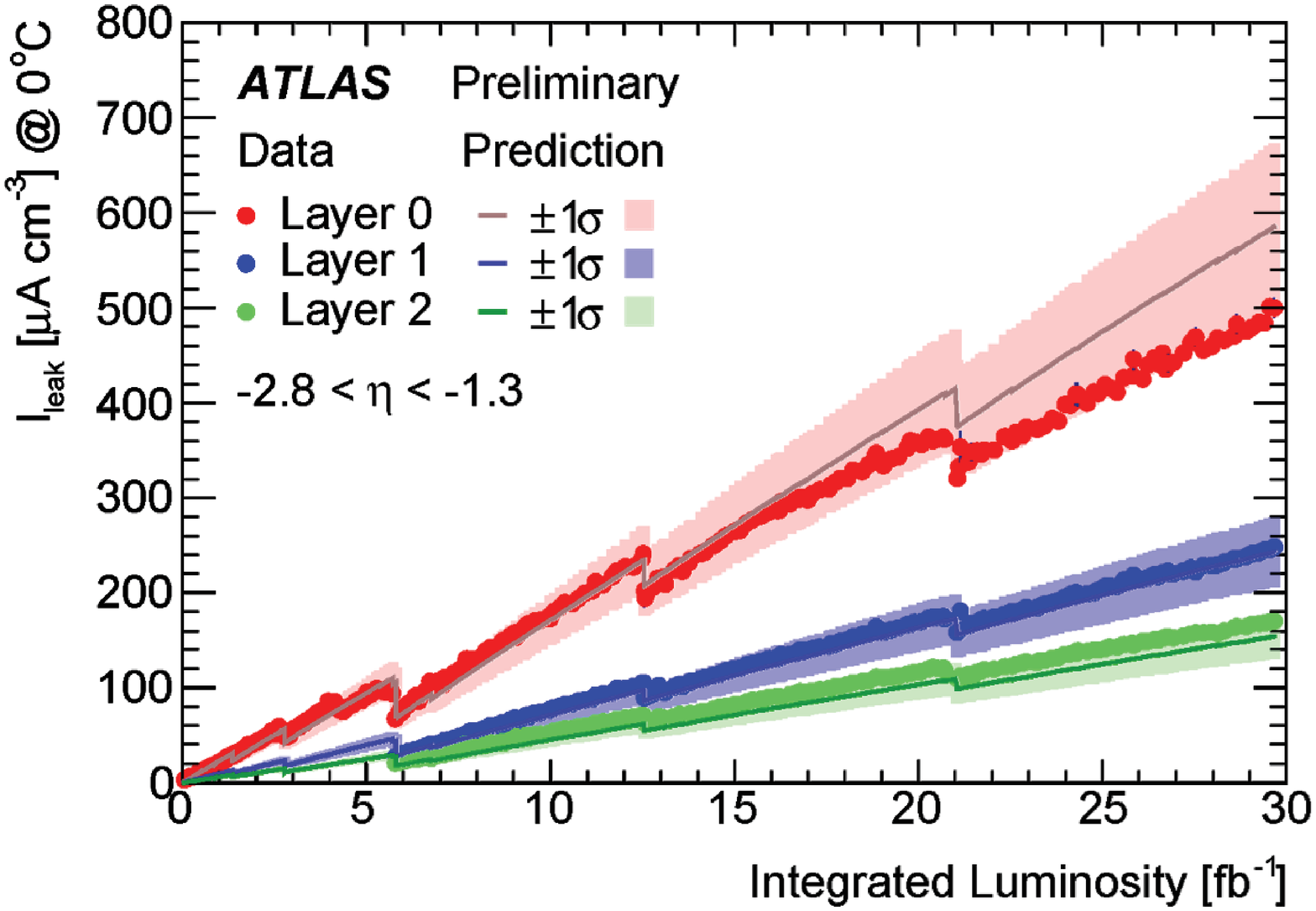}
      \caption{}
      \label{fig:eta1}
      \end{subfigure}%
      \begin{subfigure}{0.32\textwidth}
      \centering
        \includegraphics[width=1.0\linewidth]{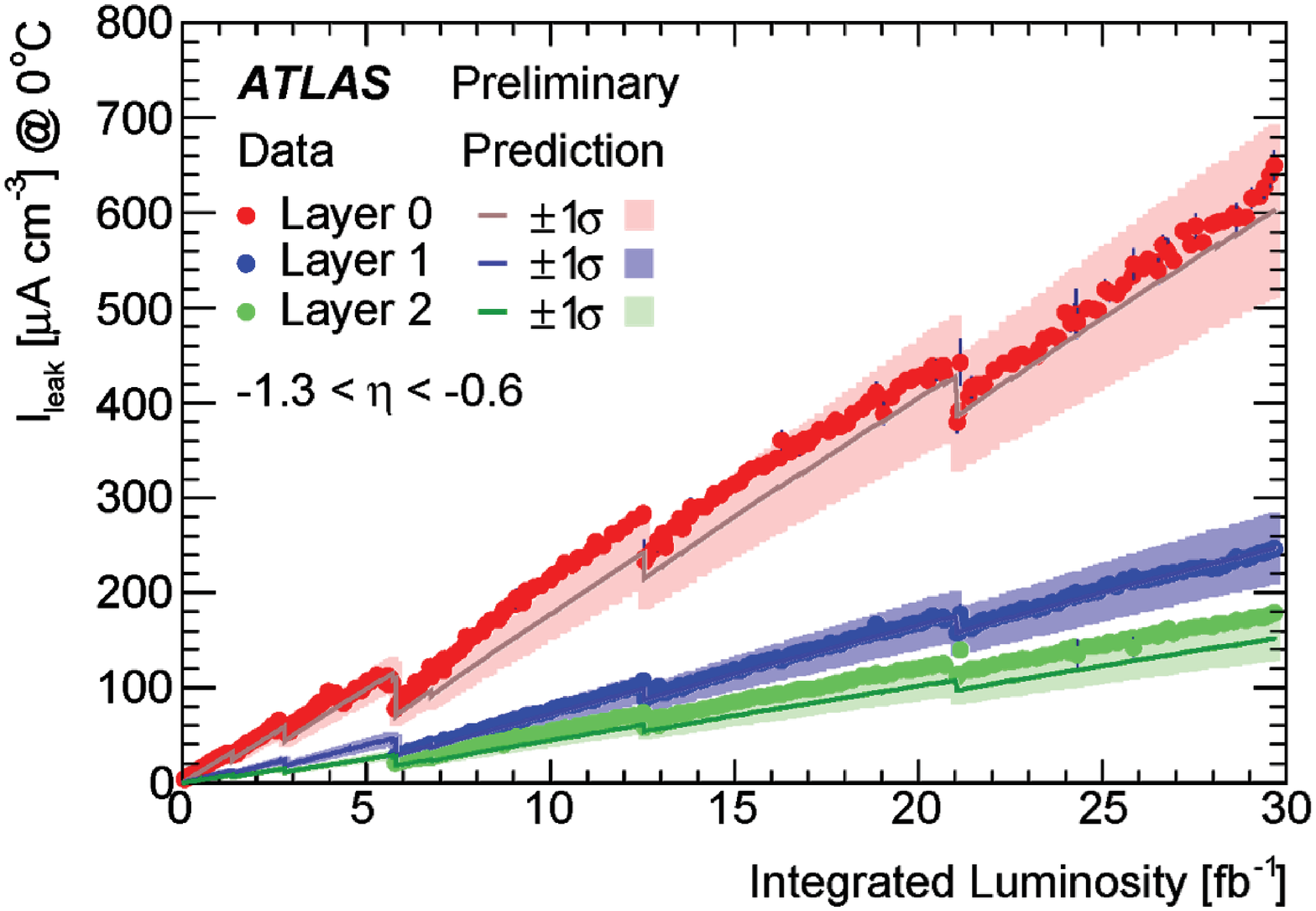}
      \caption{}
      \label{fig:eta2}
      \end{subfigure}%
      \begin{subfigure}{0.32\textwidth}
      \centering
        \includegraphics[width=1.0\linewidth]{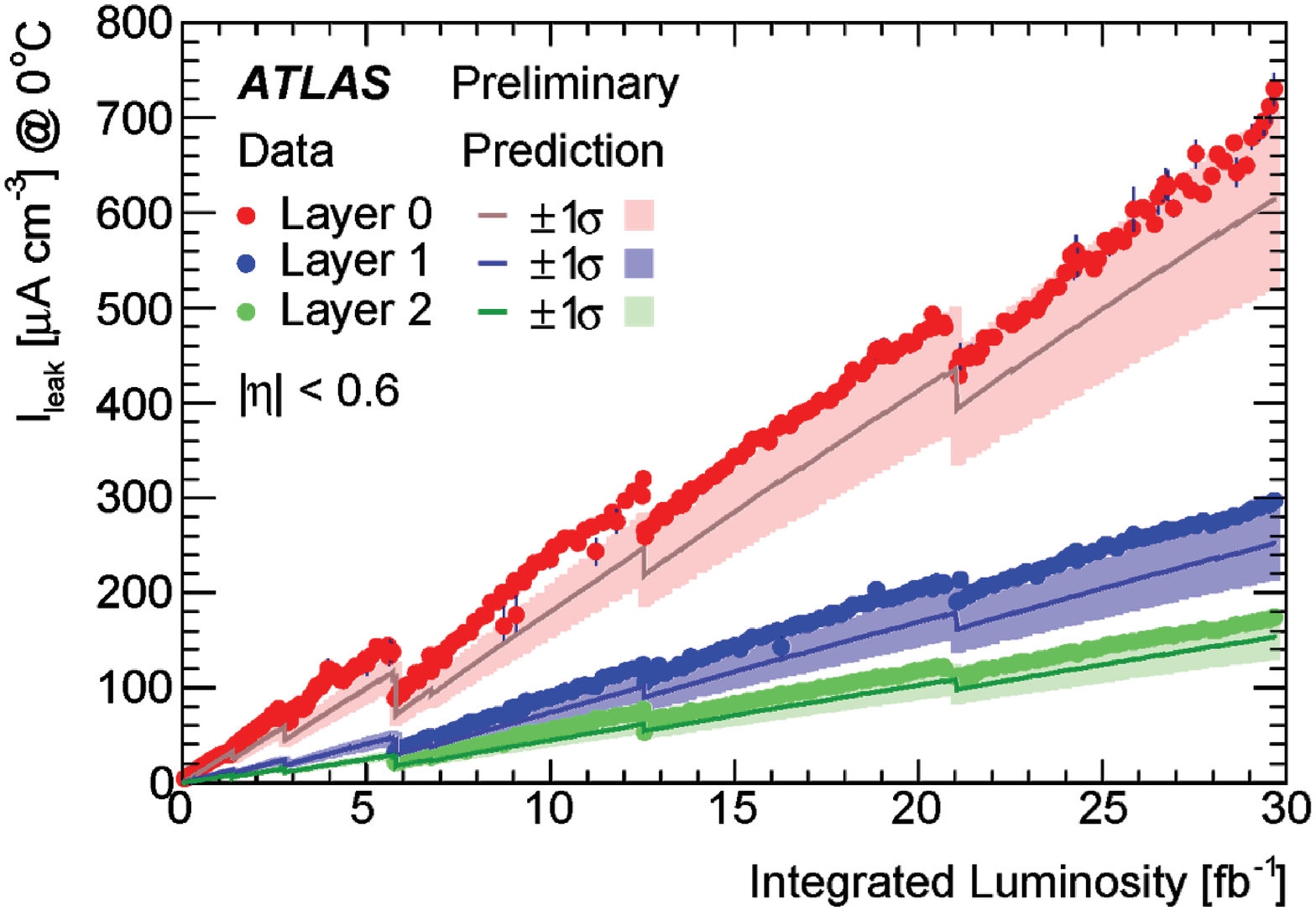}
      \caption{}
      \label{fig:eta3}
      \end{subfigure}
      \begin{subfigure}{0.32\textwidth}
      \centering
        \includegraphics[width=1.0\linewidth]{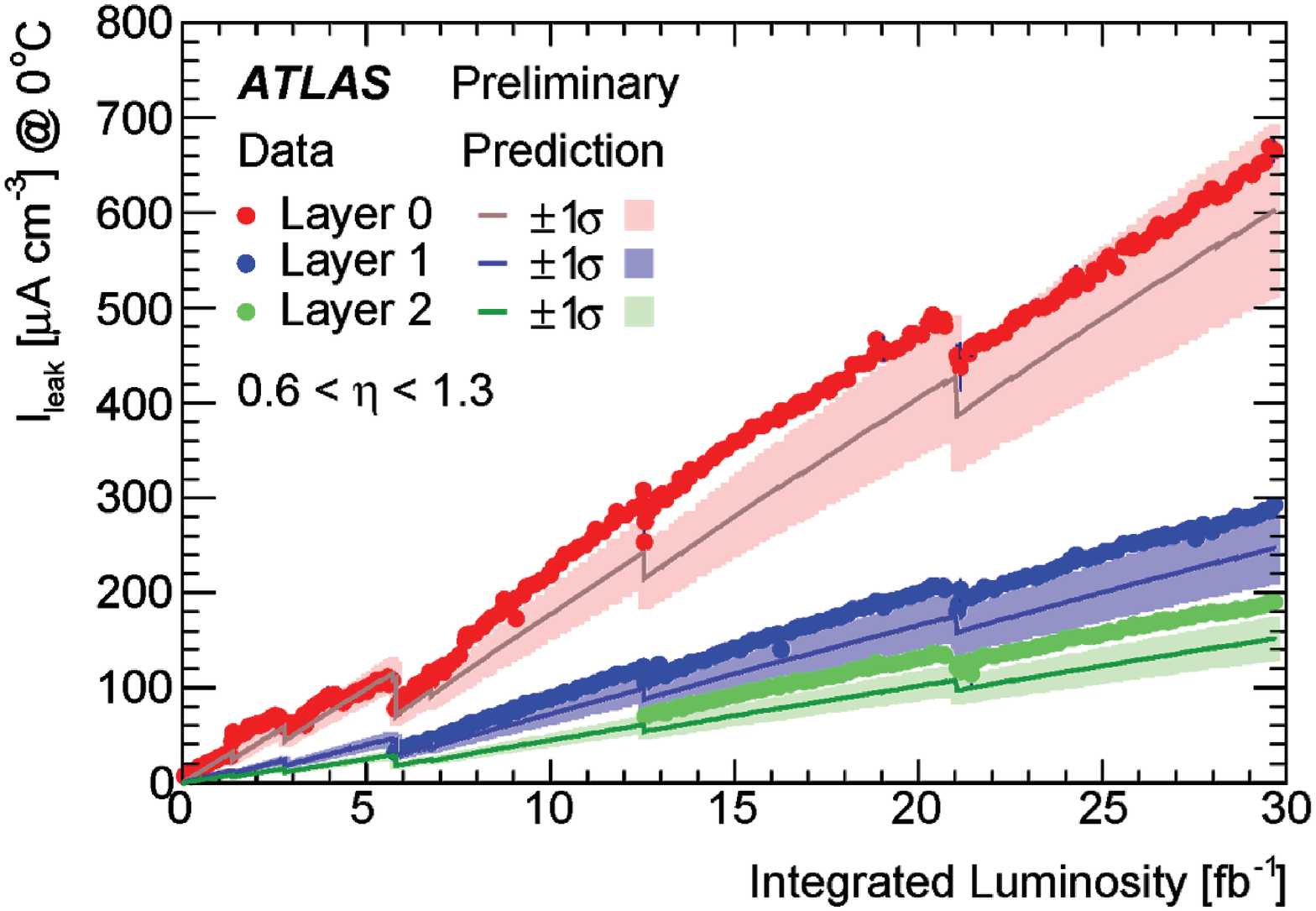}
      \caption{}
      \label{fig:eta4}
      \end{subfigure}%
      \begin{subfigure}{0.32\textwidth}
      \centering
        \includegraphics[width=1.0\linewidth]{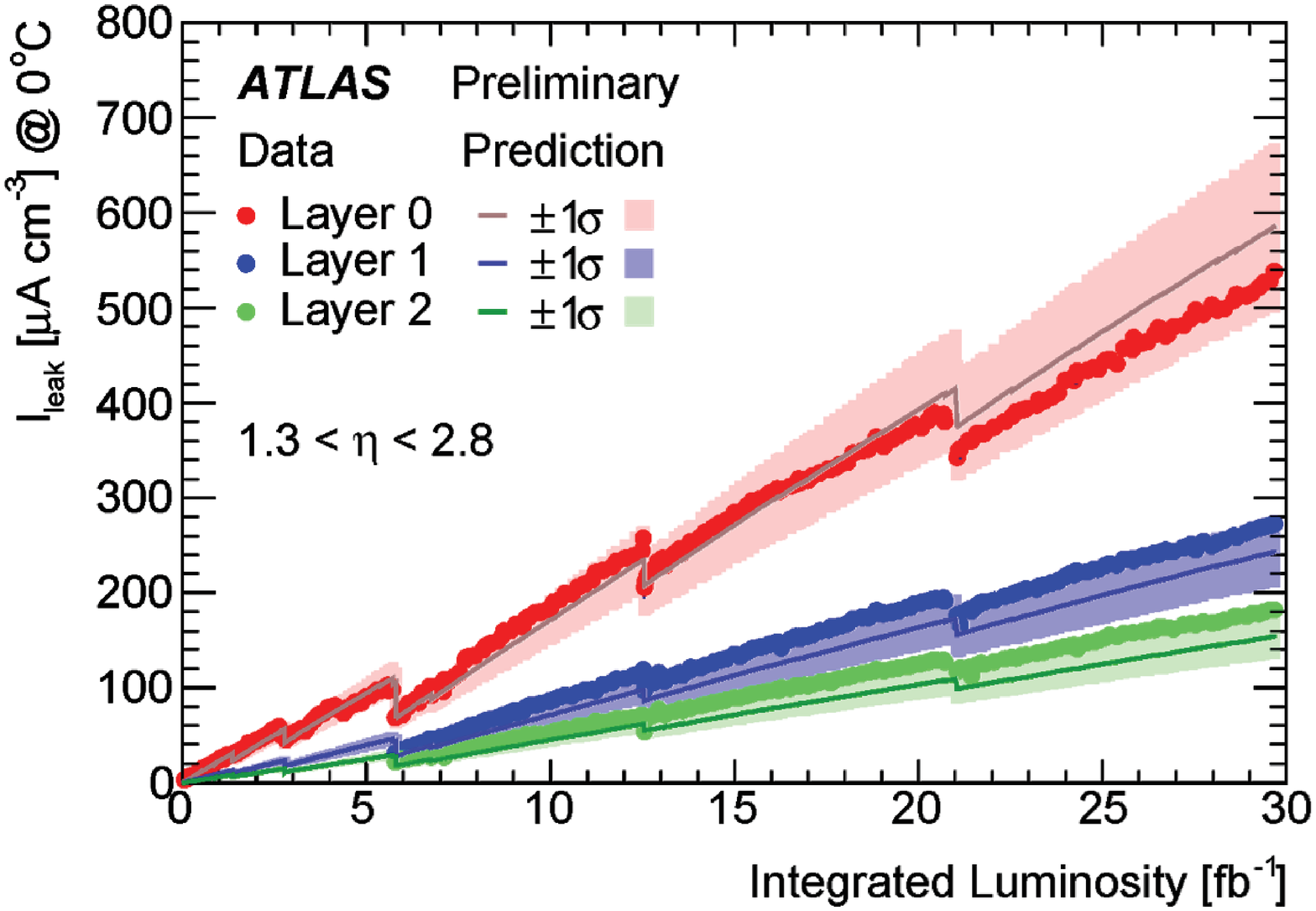}
      \caption{}
      \label{fig:eta5}
      \end{subfigure}%
      \begin{subfigure}{0.32\textwidth}
      \centering
        \includegraphics[width=1.0\linewidth]{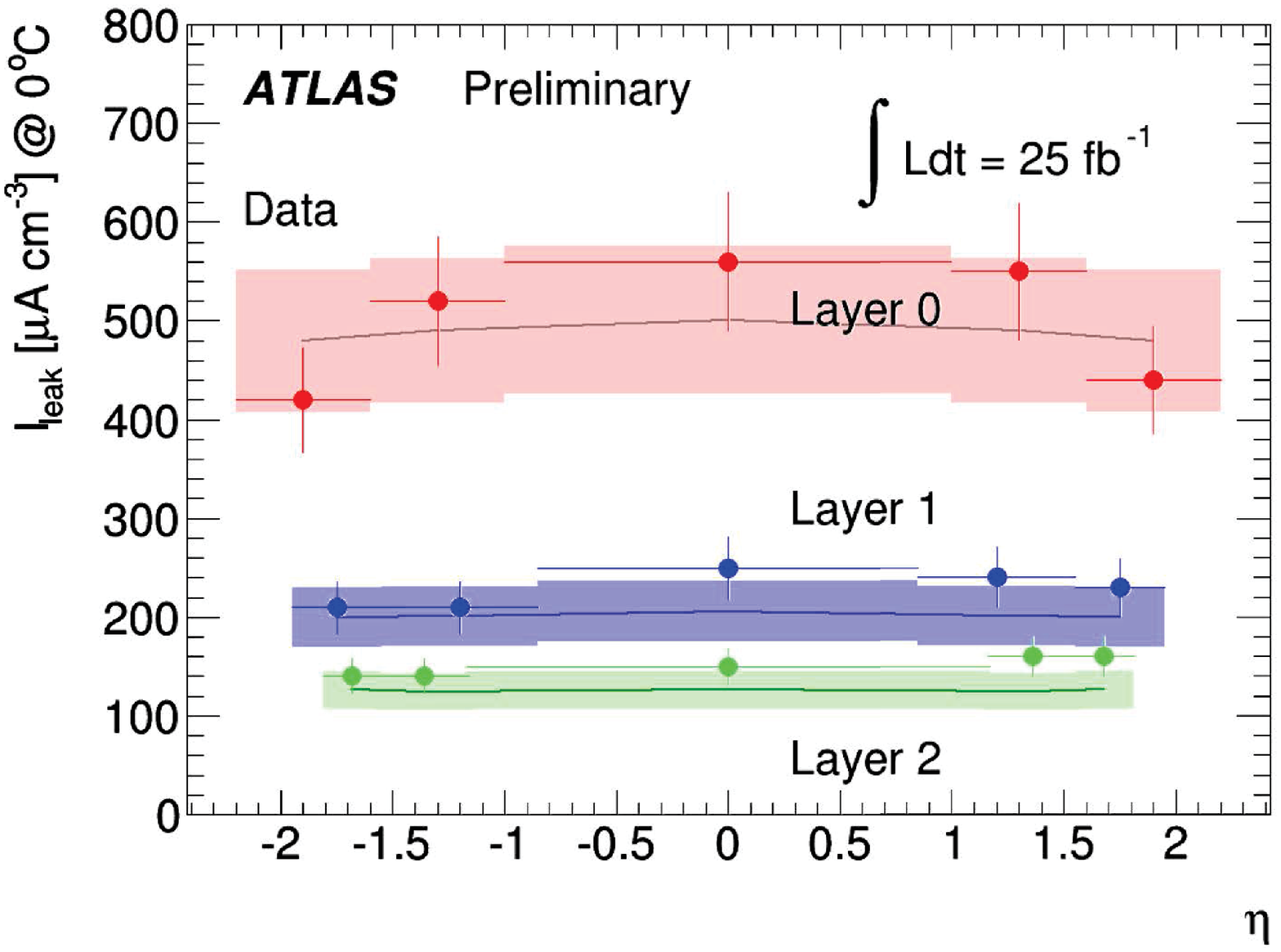}
      \caption{}
      \label{fig:eta6}
      \end{subfigure}
    \caption{ (a) through (e): ATLAS Pixel module leakage current in each
      of 5 pseudorapidity sectors, versus integrated LHC
      luminosity~\cite{ATL-INDET-PUB-2014-004}.  The currents are
      averaged over their layer for all modules equipped with Current
      Measurement Boards within the layer.  The current is continuously
      monitored by the ATLAS Detector Control System. Predictions based
      on the Hamburg Model are included.  
      Discontinuities are due to annealing during LHC and cooling stops.
      (f): the module leakage current in
      defined 5 pseudorapidity bins, for the integrated luminosity equal
      to 25\invfb~\cite{ATL-INDET-PUB-2014-004}.}
      \label{cmbpaper-eta} 
    \end{figure}
\section{Precision and Systematic Uncertainties}
  The contributions to the uncertainty on the current include
  \begin{itemize}
    \item the CMB and ELMB precision on current measurements: \(12\%\) for both ranges.
    \item the number of current measurements (two per hour).
    \item the uncertainty on the luminosity, \(1.8\%\) in 2011 and \(2.8\%\) in
          2012~\cite{Aad:2013ucp}.
    \item the temperature uncertainty, which, at less than \(0.3\degc\),
          contributes to the uncertainty on the current with \(3.4\%\).
  \end{itemize} 
\section{Lifetime Estimate}
  The temperature-corrected current readings per module can be
  extrapolated to predict the amount of current the pixel modules will
  draw after a certain integrated luminosity has been collected with the
  ATLAS pixel detector.
  As leakage current can lead to excessive power and thermal runaway,
  limits on the module lifetime \emph{due to current} can be determined
  from the limit on the bias voltage that can be applied.
  A single Iseg power supply channel can sustain a maximum current of
  \(\lsim{4000\mkamp} \) for two modules. There are other consequences
  of radiation-induced leakage current upon module lifetime, including
  the noise, bias, and threshold limitations of the front end chips as
  well as the finite capacity of the cooling system; while important,
  these are not treated here.
\par
  We extrapolate the present rate of current increase with integrated
  luminosity, assuming optimistically that the  modules will
  be exposed for 10 days to $20$\degc and otherwise maintained at
  $-13$\degc in future years.  The model assumes that the bias voltage is raised
  throughout LHC operation as needed to keep the bulk fully depleted,
  until a maximum allowed voltage of 600~V is reached.  Only the effects
  of proton runs are considered.  Figure~\ref{lifetime} shows the
  implied leakage current per module versus date and versus integrated
  luminosity for Layer-0 in this scenario~\cite{ATL-INDET-PUB-2014-004}.
  This extrapolation predicts a leakage current of about
  $500\mkamp$ per  Layer-0 module (of thickness $250\mkm$) after the accumulation of
  about 450\invfb of delivered luminosity, which corresponds to January $2024$ in the
  scenario proposed here.
\begin{figure}
\centering
  \begin{minipage}{.66\textwidth}
  \centering
      \includegraphics[width=1.0\linewidth]
      {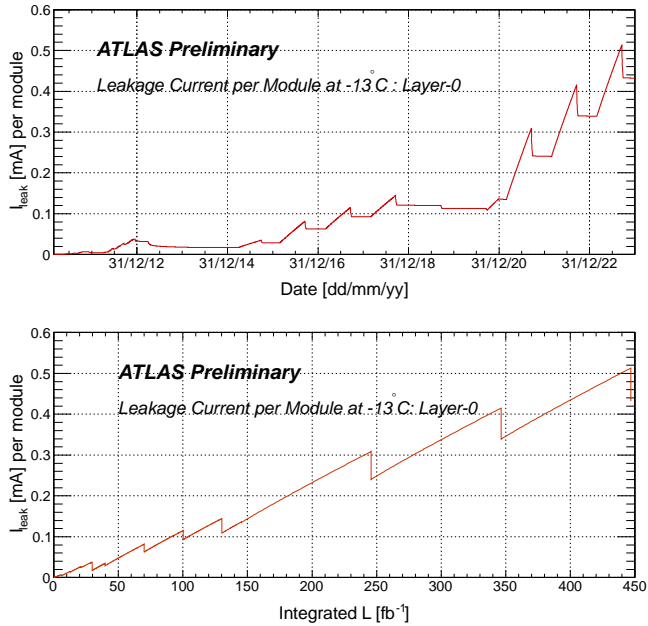}
  \end{minipage}%
  \begin{minipage}{.33\textwidth}
  \centering
    \captionof{figure}{ Predicted leakage current per Layer-0 module at
               $-13$\degc, versus date (upper) and integrated
               luminosity (lower) for the scenario in which modules will
               be exposed for 10 days to $20$\degc and otherwise
               maintained at $-13$\degc~\cite{ATL-INDET-PUB-2014-004}.} 
  \label{lifetime}              
  \end{minipage}
\end{figure}
\section{ Summary and Prospects}
\label{sec:summ} 
  We have described the principles of radiation damage monitoring using
  the current measurements provided by the circuits of the ATLAS Pixel
  high voltage delivery system.  We have observed radiation damage in
  the ATLAS Pixel Detector with characteristics in agreement with the
  Hamburg Model.  The dependence of the leakage current upon the
  integrated luminosity has been presented.  A linear behavior of the
  response has been observed.  The current distribution in azimuthal
  quadrants is symmetric as expected.  The shape of the current
  distribution in pseudorapidity is in agreement with a fluence
  calculation based on the {\sc FLUKA} transport code within the
  assigned systematic uncertainties.  Extrapolation of leakage current
  with time predicts survival of Layer-0 up to \(450\invfb\) of
  data. This current monitoring system will be expanded to encompass
  pixel disk modules in Run~2 and beyond.
\acknowledgments
  The author expresses his grateful appreciation to the colleagues from
  ATLAS Pixel Detector group for friendly cooperation and support
  through all phases of the project on pixel sensor leakage current
  monitoring and measurement.
%
%

%
%
%
\end{document}
